# A scoping review of causal methods enabling predictions under hypothetical interventions

Lijing Lin[1*], Matthew Sperrin[1], David A. Jenkins[1,2], Glen P. Martin[1], Niels Peek[1,2,3]

[1]*Division of Informatics, Imaging and Data Science, Faculty of Biology, Medicine and Health, University of Manchester, Manchester Academic Health Science Centre, Manchester, UK*

[2]*NIHR Greater Manchester Patient Safety Translational Research Centre, The University of Manchester, UK*

[3]*NIHR Manchester Biomedical Research Centre, The University of Manchester, Manchester Academic Health Science Centre, UK*

**Corresponding author:* Lijing Lin, lijing.lin@manchester.ac.uk

Background
The methods with which prediction models are usually developed mean that neither the parameters nor the predictions should be interpreted causally. For many applications this is perfectly acceptable. However, when prediction models are used to support decision making, there is often a need for predicting outcomes under hypothetical interventions.

Aims
We aimed to identify published methods for developing and validating prediction models that enable risk estimation of outcomes under hypothetical interventions, utilizing causal inference. We aimed to identify the main methodological approaches, their underlying assumptions, targeted estimands, and potential pitfalls and challenges with using the method. Finally, we aimed to highlight unresolved methodological challenges.

Methods
We systematically reviewed literature published by December 2019, considering papers in the health domain that used causal considerations to enable prediction models to be used for predictions under hypothetical interventions. We included both methodologies proposed in statistical/machine learning literature and methodologies used in applied studies.

Results
We identified 4919 papers through database searches and a further 115 papers through manual searches. Of these, 87 papers were retained for full text screening, of which 13 were selected for inclusion. We found papers from both the statistical and the machine learning literature. Most of the identified methods for causal inference from observational data were based on marginal structural models and g-estimation.

Conclusions
There exist two broad methodological approaches for allowing prediction under hypothetical intervention into clinical prediction models: 1) enriching prediction models derived from observational studies with estimated causal effects from clinical trials and meta-analyses; and 2) estimating prediction models and causal effects directly from observational data. These methods require extending to dynamic treatment regimes, and consideration of multiple interventions to operationalise a clinical decision support system. Techniques for validating ‘causal prediction models’ are still in their infancy.

## 1. Introduction

Clinical prediction models (CPMs) aim to predict current diagnostic status or future outcomes in individuals, conditional on covariates (1). In clinical practice, CPMs may inform patients and their treating physicians of the probability of a diagnosis or a future outcome, which is then used to support decision-making. For example, QRISK (2,3) computes an individual's risk of developing cardiovascular disease within the next 10 years, based on their characteristics such as BMI, blood pressure, smoking status, and other risk factors. The National Institute for Health and Care Excellence (NICE) guidelines indicate that anyone with an estimated QRISK above 10% should be considered for statin treatment (4). These guidelines also state that initially, patients should be encouraged to implement lifestyle changes such as smoking cessation and weight loss. However, guidelines using such clinical prediction models can be problematic for two main reasons. First, there is often a lack of clarity concerning the estimand that a clinical prediction model is targeting (5). To inform decision making about treatment initiation, one requires predicted risks assuming no treatment is given. This might be achieved by using a 'treatment-naïve' cohort (removing all patients who take treatment at baseline) (2,3), or by incorporating treatment as a predictor variable in the model (6). However, such approaches do not handle 'treatment drop-in': in which patients in the development cohort might start taking treatment post-baseline (7,8). One way to attempt to account for this is to censor patients at treatment initiation, however this assumes that treatment initiation is not informative (9). Second, these prediction models cannot indicate which of the potential treatment options or lifestyle changes would be best in terms of lowering an individual's future cardiovascular risk, nor can they quantify the future risk if that individual were given a treatment or lifestyle change (10,11). With the lack of randomised treatment assignment such as in the observational studies, simply 'plugging in' the hypothetical treatment or intervention via the baseline covariates will rarely, if ever, give the correct hypothetical risks (11) For example, there may be underadjustment due to residual confounding, or overadjustment of mediators or colliders.

To correctly aid such decision-making, one needs answers to 'what-if' questions. As an example, suppose we are interested in statin interventions for primary prevention of CVD and we would like to predict the 10-year risk of CVD with or without statin interventions at an individual level. The methods used to derive CPMs do not allow for the correct use of the model in answering such 'what-if' questions, as they select and combine covariates to optimize predictive accuracy, not to predict the outcome distribution under hypothetical interventions (12,13). Nevertheless, end-users often mistakenly compare the contribution of individual covariates (in terms of risk predictions) and seek causal interpretation of model parameters (14). Within a potential outcomes (counterfactual) framework, an emerging class of *causal predictive models* could enable 'what-if' queries to be addressed, specifically calculating the predicted risk under different hypothetical interventions. This enables targeted intervention, allows correct communication to patients and clinicians, and facilitates a preventative healthcare system.

There exists a vast literature on both predictive models and causal inference. While the use of prediction modelling to enrich causal inference is becoming widespread (15), the use of causal thinking to improve prediction modelling is less well studied (16), however its potential is acknowledged (13). Our aim was therefore to identify methods for developing and validating 'causal prediction' models that use causal methods to enable risk estimation of outcomes under hypothetical interventions. We aimed to identify the main methodological approaches, their underlying assumptions, targeted estimands, and possible sources of bias. Finally, we aimed to highlight unresolved methodological challenges.

## 2. Methods

We aimed to identify all studies in which a form of causal reasoning is used to enable predictions for health outcomes under hypothetical interventions. To be clear, we were not interested in causal studies where the methods can solely be used to predict average or conditional causal effects (13).

Due to the available resources for reviewing large volumes of papers, the search was restricted to the health domain. We included both methodologies proposed in statistical/machine learning literature and methodologies used in applied studies. The review process adhered to Arksey and O'Malley's (17) scoping review framework and the preferred reporting items for systematic reviews and meta-analyses (PRISMA) statement (18). We have also followed recommendations for conducting methodology scoping reviews as suggested in Martin *et al.* (19).

### 2.1 Search strategy

We systematically reviewed the literature available up to a cut-off date of 31 December 2019. The literature search was conducted in two electronic databases: Ovid Medline and Ovid Embase, and searches were tailored to each database and restricted to English language publications. The search terms were designed by considering the intersection of prediction modelling and causal inference. Pre-existing search filters were utilised where possible such as those for prediction models (20). Details of the search terms are included in the supplementary protocol. We were also aware, *a priori,* of several research groups that have published work on methods in related areas (listed in the supplementary protocol). We manually searched for any relevant recent publications within the past 4 years from these groups. In addition, we conducted backward citation search checking the references of identified papers, and a forward citation search using Google Scholar, which discovered papers referencing the identified papers.

### 2.2 Selection of studies

After the initial search, all titles and abstracts of papers identified by the search strategy were screened for eligibility by the lead author (LL). A random 3% were screened by a second reviewer (DJ) to ensure reliability of the screening process. Any discrepancies between the reviewers were solved through mutual discussion, in consultation with a third reviewer, where needed (MS). The initial eligibility criteria, based on title and abstract screen, were as follows: (1) use causal reasoning in the context of health outcome prediction, specifically enabling prediction under hypothetical interventions; (2) describe original methodological research (e.g. peer reviewed methodological journal); or (3) applied research, which did not develop methodology, but state-of-the-art methodology was employed to address relevant causal prediction questions. We excluded studies that could only be used for causal effect estimation, and studies where standard clinical prediction models were used to infer conditional causal effects, e.g. (21). However, we do not exclude papers that developed a novel method of allowing prediction under hypothetical intervention, even when the final goal was causal effect estimation. We excluded letters, commentaries, editorials, and conference abstracts with no information to allow assessment of proposed methods.

### 2.3 Extraction

Following the review aims, we extracted information from papers that were included after full-text screening as follows:

1. Article type (summary/review, theoretical, modelling with application via simulation and/or observed data, purely applied paper);

2. Clinical topic area of analysis (e.g. CVD, HIV, cancer) for papers with application to observed data;

3. Intervention scenarios (single intervention vs multiple interventions); types of outcome and exposure outcomes examined (binary, time-to-event, count, continuous, other);

4. Information on targeted estimand and possible validation approaches for the proposed methods inferred by the review authors; stated possible sources of bias;

5. Stated assumptions; methodologies/methods used for the causal effect estimation and outcome prediction; main methodological novelty stated by the authors of identified papers;

6. Reported modelling strengths and limitations, and suggestions for the future work;

7. Availability of software/code.

The completed extraction table is available in the Supplementary File 1. Categorisation of papers was carried out during information extraction phase by synthesising the extracted information.

## 3. Results

Our database searches identified 4919 papers. We identified a further 115 papers through checking publications from known research groups, and forward and backward citation searching. Of these, 87 were retained for full text screening, with 13 of these were deemed eligible for final inclusion, as listed in Table 1. The process of study identification, screening and inclusion is summarised in the PRISMA flowchart (Figure 1).

The identified papers covered two main intervention scenarios: single intervention (22–27) and repeated interventions over time (8,28–33) with nearly an equal amount of papers addressing average intervention effects (defined by a contrast of means of counterfactual outcomes averaged across the population) (8,22,23,25,29,30) and conditional effects (defined as the contrast of covariate-specific means of the outcome under different intervention levels) (24,26–28,31–33). Across the included papers, we identified two broad categories of methodological approaches for developing causal prediction models: (1) enriching prediction models with externally estimated causal effects, such as from meta-analyses of clinical trials; and (2) estimating both the prediction model and causal effects from observational data. The majority of the identified papers (10 out of 13) fell into the latter category, which can be further divided according to intervention scenarios and included methods embedded within both statistical and machine learning frameworks. Table 1 describes part of the extracted information on each paper. The complete extraction table is available in the Supplementary File 1. In addition, we will illustrate the methods identified using the statin interventions for primary prevention of CVD example introduced previously, explaining each method and showing their differences in terms of targeted questions and corresponding estimand (Table 2).

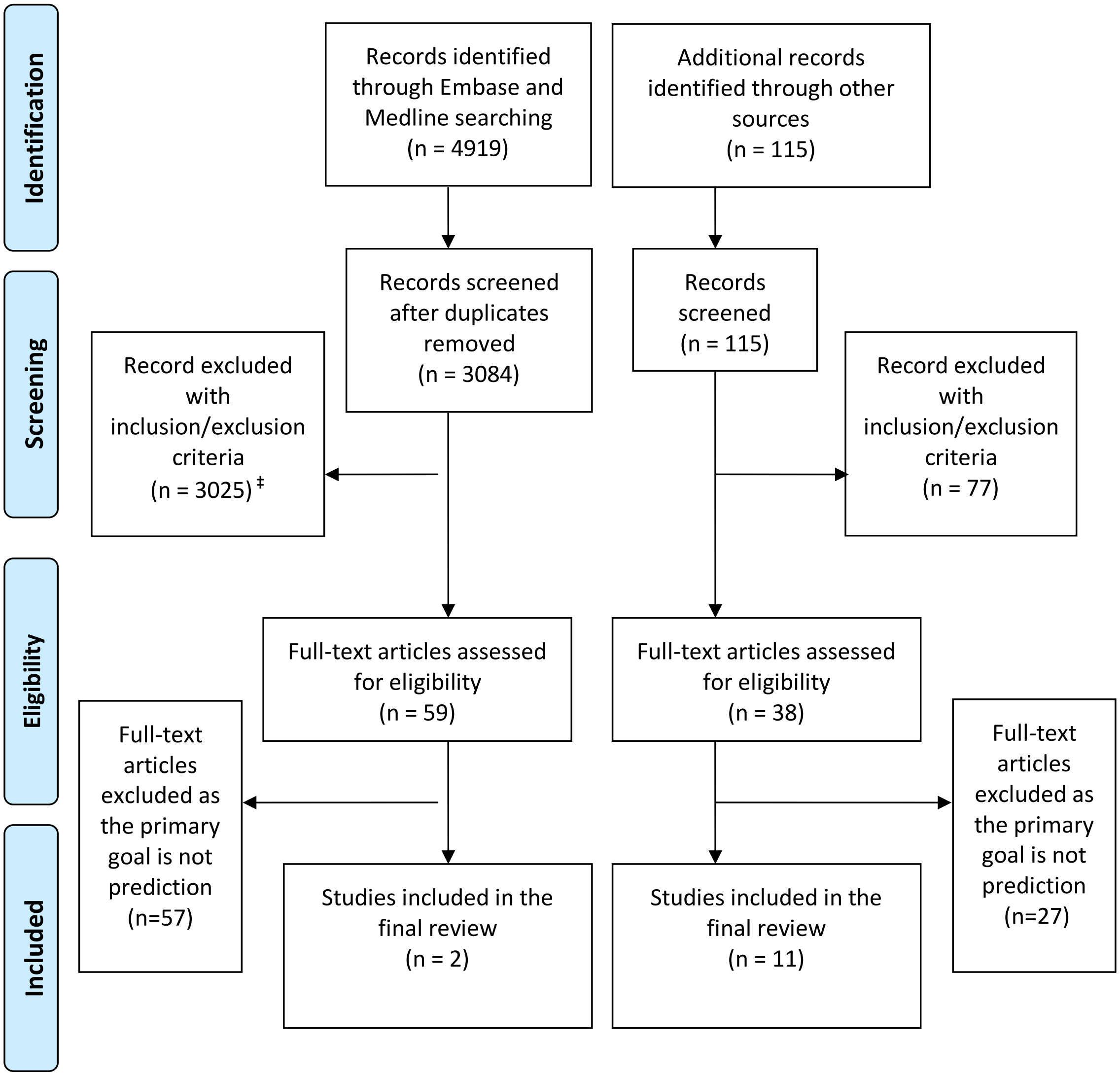

Figure 1. PRISMA flow diagram. ‡ The majority of papers excluded at this stage did not meet our first inclusion criterion: that is, they did not focus on prediction under hypothetical interventions; these papers either did not use prediction models at all, or only used prediction modelling to enrich causal inference.

Table 1. Summary of included 13 papers (See Supplementary File 1 for the completed extraction table). Abbreviations for 'Stated Assumptions': A(1): Relevant directed acyclic graphs (DAGs) available; A(2): Identifiability conditions (consistency, exchangeability, and positivity; or sequential version of consistency, exchangeability, and positivity for time-varying treatments); A(3): Continuous-time exchangeability; A(4): Non-informative measurement times. Other abbreviations: EHR: electronic health record; RCT: randomised controlled trial.

| Title | Intervention Scenario | Clinical topic area | Types of outcomes | Stated assumptions | Reported limitations | Code availability |
|---|---|---|---|---|---|---|
| **Candido dos Reis, F. J. et al. (2017) An updated PREDICT breast cancer prognostication and treatment benefit prediction model with independent validation, Breast Cancer Research, 19(1), 58** | Single intervention, Discrete choice, Average effect | Breast cancer | Survival | Generalisability of effect from clinical trial. | Prediction of non-breast cancer deaths was excellent in the model development data set but could under-predict or over-predict in the validation data sets. | Stata code are available from the author on request. |
| **Brunner, F. J. et al. (2019) Application of non-HDL cholesterol for population-based cardio-vascular risk stratification: results from the Multinational Cardiovascular Risk Consortium., The Lancet 394.10215: 2173-2183.** | Single intervention, Discrete choice, Average effect | CVD | Binary | The therapeutic benefit of lipid-lowering intervention investigated in the study is based on a hypothetical model that assumes a stable reduction of non-HDL cholesterol. | (1) Data limitation in the derivation cohort. (2) Strong clinical assumption that treatment effects are sustained over a much longer term than has been studied in clinical trials. | Reported using R but codes not available |
| **Silva, R. (2016), Observational-Interventional Priors for Dose-Response Learning. In Advances in Neural Information Processing Systems 29.** | Single intervention, Treatment dose (continuous), Conditional effect | Infant Health and Development Program (IHDP) | Continuous | A(1) and additionally: It is possible to collect interventional data such that treatments are controlled | (1) Computation complexity. (2) Have not discussed at all the important issue of sample selection bias. (3) Generalisability issue. | Code available from OLS |
| **Van Amsterdam, W. A. C. et al. (2019). Eliminating biasing signals in lung cancer images for prognosis predictions with deep learning. npj Digital Medicine, 2(1), 1-6.** | Single intervention, Discrete choice (0/1), Average effect | Lung caner | Survival | A(1), A(2) and additionally: An image is hypothesized to contain important information for the clinical prediction task. The collider can be measured from the image. | (1) Provide an example of how deep learning and structural causal models can be combined. Methods combining machine learning with causal inference need to be further developed. | Code available from OLS |
| **Alaa, A. M., & Van Der Schaar, M. (2017). Bayesian Inference of Individualized Treatment Effects using Multi-task Gaussian Processes. In Advances in Neural Information Processing Systems 30.** | Single intervention, Discrete choice (0/1), Conditional effect | IHDP & Heart transplantation for cardiovascular patients | Continuous/Survival times | A(2) | (2) No experiments regarding outcome prediction accuracy. (2) The computational burden is dominated by the O(n3) (matrix inversion on line 13 in Alg.1. | Code available from authors' website. |
| **Arjas, E. (2014) Time to Consider Time, and Time to Predict? Statistics in Biosciences. Springer New York LLC, 6(2), pp. 189-203** | Single intervention, Discrete choice, Conditional effect | Acute middle ear infections | Survival | A(2) and local independence | In studies involving real data the computational challenge can become formidable and even exceed what is feasible in practice. | NA |
| **Pajouheshnia R. et al. (2020) Accounting for time-dependent treatment use when developing a prognostic model from observational data: A review of methods. Stat Neerl. 74(1).** | Multiple intervention; Discrete choice (0/1); Average effect | Chronic obstructive pulmonary disease (COPD) | Survival | A(1) and A(2) | A very strong indication for treatment will result in structural non-positivity leading to biased estimates of treatment-naïve risk. | NA |
| **Sperrin, M. et al. (2018) Using marginal structural models to adjust for treatment drop-in when developing clinical prediction models, Statistics in Medicine. John Wiley & Sons, Ltd, 37(28), pp. 4142-4154.** | Multiple intervention; Discrete choice (0/1); Conditional effect | CVD | Binary | A(1) and A(2) | (1) Have not modelled statistical interaction between treatment and prognostic factors; (2) Did not explicitly model statin discontinuation; (3) Only consider single treatment. | Code available from OLS |
| **Lim, B. et al. (2018). Forecasting Treatment Responses Over Time Using Recurrent Marginal Structural Networks. In Conference on Neural Information Processing Systems 32.** | Multiple intervention; No restriction on treatment choices; Average effect | Cancer growth and treatment responses | No restriction | A(2) | NA | Code available from OLS |
| **Bica, I. et al. Estimating Counterfactual Treatment Outcomes over Time through Adversarially Balanced Representations. ICLR 2020** | Multiple intervention; Discrete treatment choices; Average effect | Treatment response in a tumour growth model | No restriction | A(2) | Additional theoretical understanding is needed for performing model selection in the causal inference setting with time-dependent treatments and confounders. | Code available from authors' website. |
| **Xu, Y. et al. (2016) A Bayesian Nonparametric Approach for Estimating Individualized Treatment-Response Curves. Edited by F.Doshi-Velez et al. PMLR , pp. 282-300.** | Multiple intervention; Discrete treatment choices; Conditional treatment effect | (1) kidney function deterioration in ICU; (2) the effects of diuretics on fluid balance. | Continuous | A(2) | NA | NA |
| **Soleimani, H. et al. (2017). Treatment-response models for counterfactual reasoning with continuous-time, continuous-valued interventions. In Uncertainty in Artificial Intelligence. Proceedings of the 33rd Conference, UAI 2017.** | Continuous-time intervention; Continuous-valued treatments; Conditional treatment effect | Modelling physiologic signals with EHRs for treatment effects on renal function | No restriction | A(2), A(3) | While this approach relies on regularisation to decompose the observed data into shared and signal-specific components, new methods are needed for constraining the model in order to guarantee posterior consistency of the sub-components of this model. | NA |
| **Schulam, P., & Saria, S. (2017). Reliable Decision Support using Counterfactual Models. In Advances in Neural Information Processing Systems 30.** | Continuous-time intervention; Continuous-valued treatments; Conditional treatment effect | Applicable to data from EHR but not restrict to such medical settings | Continuous-time; no restriction on data type | A(2), A(3), A(4) | (1) the validity of the CGP is conditioned upon a set of assumptions that are, in general, not testable. The reliability of approaches therefore critically depends on the plausibility of those assumptions in light of domain knowledge. | Code available from authors' website. |

Table 2. Illustration of methods in different categories using an example of statin intervention in primary prevention of CVD.

| Approach categories | | | Refs | Targeted estimand | Potential pitfalls/challenges | Exemplary methods/Evaluations |
|---|---|---|---|---|---|---|
| Combining causal effects measured from external information | Two-stage approach | | Candido dos Reis et al. (22) | Risk of CVD under intervention of taking or not taking statin at baseline (and, in a considered trial protocol, following-up for a certain length of time during which statin choice is maintained): $E\left(Y^{(A_0)} \mid X_0\right)$ | Efficacy/effectiveness gap when translating trial results to routine care. Comparability of trail and observed populations (selection bias). | Develop a CPM using individuals who take statin at baseline with the coefficient for treatment variable in the model fixed to the statin effects estimated from trials. |
| | | | Brunner et al. (23) | | Inflating the baseline cholesterol for individuals receiving statin by a certain level has assumed that ‘statins had a moderate effect on lipid reduction and was initiated late during lifetime’, and that statins operate only through cholesterol, i.e. ignores any other causal pathways. | Inflate the baseline cholesterol of individuals receiving statin (by 30% e.g.). Develop a CPM using all individuals. Combine the predicted individual-level CVD risk with an effect equation estimated from trials to get the absolute risk under intervention. |
| | One-stage approach | | Silva (24) | Risk of CVD under intervention of taking statin of dosage $a_i, (i = 1, \ldots, d)$ at baseline: - $E\left(Y^{(A_0=a_i)} \mid X_0\right)$. | Sample selection bias between the interventional data and observational data. | Individual patient data from RCTs and observational clinical data are combined under a Bayesian framework to predict risk under intervention. Use MCMC to approximate the posterior distributions of the parameters in the model. |
| Estimating both a prediction model and causal effects from observation al data | Single intervention | Related to average treatment effect estimation | Van Amsterda m et al. (25) | Risk of CVD under intervention of taking/not taking statin at baseline, regardless of future: $E\left(Y^{(A_0)} \mid X_0\right)$. | An over-simplified causal structure can lead to biased estimates of causal effects, e.g. when there exists more than one collider that were not observed but whose information were contained in the prognostic factors. | Use a CNN to separate the unobserved collider information from other risk factors while using the last layer resembling linear regression to include the treatment variable as a covariate for risk prediction under intervention. |
| | | Related to conditional treatment effect estimation | Alaa et al. (26) | Risk of CVD under intervention of taking/not taking statin at baseline, regardless of future: $E\left(Y^{(A_0)} \mid X_0\right)$. | Without careful examination of causal structure within the variables, biased association between treatment and outcome can be introduced. | Estimate the outcome curves for the treated samples and untreated samples simultaneously using the signal-in-white-noise model. The estimation of model is done through one loss function, known as the precision in estimating heterogeneous effects (PEHE). |
| | | | Arjas (27) | Risk of CVD under intervention of taking/not taking statin at baseline, regardless of future: $E\left(Y^{(A_0)} \mid \bar{H}_0\right)$. | Potentially biased estimate due to misspecification of intensity functions required in the outcome hazard model. | Use treatment history and other risk factors measured over-time to set up a Bayesian model to estimate the outcome risk intensity function over time. For prediction, given an individual’s measurements up to time $t$, estimate the risk under a single intervention by applying MCMC on the predictive distributions. |
| | Time-dependent treatments and treatment- | MSMs within a prediction model framework | Pajouhesh nia et al. (8) | Risk of CVD under interventions of taking/not taking statin at baseline and/or some other times at the future: $E\left(Y^{(\bar{A}(0,K)=0)} \mid X_0\right)$ | The effectiveness of bias correction depends on a correct specification of treatment model. | Assume a causal structure. Estimate treatment censoring probabilities by fitting logistic regression models in each of the follow-up periods and derive time-varying censoring weights. After censoring, develop the prognostic model using a weighted Cox model. |
| | confounder feedback | | Sperrin et al. (28) | Risk of CVD under interventions of taking/not taking statin at baseline and/or some other times at the future: $E\left(Y^{(\bar{A}(0,K)=0)}\middle\|X_0\right)$ | The effectiveness of bias correction depends on a correct specification of treatment model. Requires agreement between the prediction model and the set of variables required for conditional exchangeability. | Assume a causal structure. Collect the baseline prognostic factors, treatments, and treatment confounders at each time point post-baseline. Compute IPTWs using a treatment model; with derived IPTWs, build a logistic regression for outcome prediction under treatments. |
| | | | Lim et al. (29) | Risk of CVD and/or other outcomes of interest (e.g. cholesterol, SBT, etc) under multiple interventions planned for the next $\tau$ timesteps from current time, given an observed history $\bar{H}_0$: $E(Y_\tau^{(\bar{A}(0,\tau-1))}\|\bar{H}_0)$. | Requires agreement between the prediction model and the set of variables required for conditional exchangeability. | With observed treatment, covariate and outcome histories (allowing for multiple treatment options of different forms), develop a propensity network to compute the IPTW and a sequence-to-sequence model that predict the outcome under a planned sequence of interventions. |
| | | Methods based on balanced representation approach | Bica et al. (30) | | Potential confounders as no careful examination of causal structure. | Build a counterfactual recurrent network to predict outcomes under interventions:<br>1. For the encoder network, use an RNN, with LSTM unit to build treatment invariant representations of the patient history $\Phi(\bar{H}_t)$ and to predict one-step-ahead outcomes $Y_{t+1}$;<br>2. For the decoder network, use $\Phi(\bar{H}_t)$ to initialize the state of an RNN that predicts the counterfactual outcomes for future treatments. |
| | | Methods with g-computation for correcting time-varying confounding | Xu et al. (31) | Cholesterol or other continuous outcome of interest (univariate) at any time $t$ in the future, under a sequence of interventions planned irregularly from current time till $t$, $\bar{A}_{0,<t}$, given observed history: $E(Y_t^{(\bar{A}_{0,<t})}\|\bar{H}_0)$. | Potential bias due to strong assumptions on model structure and possible model misspecification. | With observed treatment/covariate/outcome histories, estimate treatment-response trajectories using a Bayesian nonparametric or semi-parametric approach:<br>1. Specify models for different components in the generalised mix-effect model for outcome prediction.<br>These usually include: treatment response, baseline regression (fixed effects), and random effects. For the case where the treatments are continuously-administrated, model the treatment response using LTI dynamic systems (Soleimani et al).<br>2. Choose priors for these models based on expert domain knowledge.<br>3. Use maximum a posteriori (MAP) (Soleimani et al.) or MCMC (Xu et al.) to approximate the posterior distributions of the parameters in the proposed model. |
| | | | Soleimani et al. (32) | $E\left(Y_t^{(\bar{A}_{0,<t})}\middle\|\bar{H}_0\right)$: same as in the Xu et al. method except that now the outcome $Y$ can be multivariate (e.g. simultaneously predict risk of CVD, cholesterol and SBT) and the treatment can be both discrete-time and continuous-time. | | |
| | | | Schulam et al. (33) | $E(Y_t^{(\bar{A}_{0,<t})}\|\bar{Y}_0, \bar{A}_{<0})$: same as in the Soleimani et al. method except that the observations only include intervention and outcome histories. | Potential bias due to strong assumptions on model structure and possible model misspecification. Lack of effect heterogeneity due to omitting baseline covariates. | With observed histories, jointly model intervention and outcomes using a marked point process (MPP):<br>1. Specify models for the components in the MPP intensity function: event model, outcome model, action (intervention) model.<br>The parameterization of the event and action models can be chosen to reflect domain knowledge. The outcome model is parameterized using a GP. |
| | | | | | | 2. Maximise the likelihood of observational traces over a fixed interval to estimate the parameters. |

### 3.1 Combining causal effects measured from external information

Three papers (22–24) were identified as developing models with combined information from different sources to address single treatment effect. Candido dos Reis *et al.* (22) and Brunner *et al.* (23) took a two-stage approach, in which treatment effect estimates from external sources such as RCTs and meta-analyses were first identified, then combined with prediction models to allow predictions under treatment. In the statins for CVD example, the method proposed by Candido dos Reis *et al.* (22) corresponds to developing a CPM including individuals who take statin at baseline, where the coefficient for the statins variable in the model is fixed to the statin effects estimated from trials. Brunner *et al.* (23) developed a CPM for cardiovascular risk which was then combined with an externally estimated equation of proportional risk reduction per unit LDL cholesterol reduction to aid decision making in lipid-lowering treatment usage.

In addition to the above two-stage approach borrowing causal information estimated externally into predictive models, a one-stage approach, proposed by Silva (24), was also identified where the two sources of data, interventional and observational, were jointly modelled for causal prediction. This approach was applied in a scenario where it is possible to collect interventional data such that treatments were controlled but where sample sizes might be limited. The idea was to transform observational data into informed priors under a Bayesian framework to predict the unbiased *dose-response curve* under a pre-defined set of interventions, or 'dose'.

All the three approaches above are limited to a single intervention type and intervening at a single point in time, where, in a considered trial protocol, the intervention may follow-up for a certain length of time during which its choice is maintained (e.g. the initialisation of statin intervention). Approaches that directly apply the externally estimated causal effects into CPMs assume that the estimated causal effects are generalisable to the population in which one wishes to apply the prediction model. Equally, combining individual data from both sources (i.e. the one-stage approach) ignored the issue of sample selection bias, which was highlighted in (24). Additionally, the one-stage approach can become computationally intensive as the size of the observational and number of treatment levels increase.

### 3.2 Estimating both a prediction model and causal effects from observational data

A total of 9 papers discussed modelling predictions under interventions entirely from observational data. Approaches from these papers can be further divided into two categories: (1) methods considering only one intervention at a single time point (25–27), as discussed in the following section 3.2.1, and (2) methods allowing time-dependent interventions (8,28–33), as discussed in section 3.2.2.

### 3.2.1 Counterfactual prediction models that consider an intervention at a single point in time

In our running example, this corresponds to a decision at a single time of whether to prescribe statins for CVD prevention. It does not account for whether statins are discontinued or started at any subsequent time.

***Related to average treatment effect estimation***

Decision-making on whether to intervene on treatment requires an unbiased estimate of the treatment effect at baseline. Assuming that the Directed Acyclic Graph (DAG) that encodes the relationship between all the relevant variables is known, then *do-calculus* (34) provides an indication of whether this can be achieved in the setting of observational data with the required causal assumptions. For example, including a collider in the model (a variable caused by both treatment and outcome) will lead to biased estimates of treatment effects on the outcome. A more complex scenario appears when the collider itself cannot be directly observed but its information is contained in other prognostic factors. Van Amsterdam *et al.* (25) proposed a deep learning framework to address this particular scenario. Their goal is to predict survival of lung cancer patients using CT-scan images, in which case factors such as tumor size and heterogeneity are colliders that cannot be directly observed but can be measured from the image. The authors proposed a multi-task prediction scheme embedded in a convolutional neural network (CNN) framework (a non-linear model often used with images) which can simultaneously estimate the outcome and the collider. It used a CNN to separate the unobserved collider information from images while enabling the treatment to be appropriately included as a covariate for risk prediction under intervention. As there is no modelling of interactions between treatment and other covariates, this approach only addresses average treatment effects.

Van Amsterdam *et al.* (25) has demonstrated that deep learning can in principal be combined with insights from causal inference to estimate unbiased treatment effect for prediction. However, the causal structure applied therein was in its simplest form, and further developments are needed for more realistic clinical scenarios where, e.g., there is confounding for treatment assignment, or a treatment effect modifier exists within the image.

***Related to conditional treatment effect estimation***

Let $Y^{(a)}$ denote the potential outcome under an intervention $a$. For example, one's risk of CVD or cholesterol level under intervention of taking statin. Conditional treatment effects for subjects with a covariate $X = x$ in a population at a single time point is defined as $T(x) = E[Y^{(1)} - Y^{(0)}|X = x]$ and our goal here is to estimate the counterfactual prediction of $E\left[Y^{(a)}\middle|X = x\right], a \in \{0,1\}$. In an RCT, given complete randomisation – i.e. $a$ is independent of $Y^{(a)}$ and $X$, under consistency, one can estimate $E\left[Y^{(a)}\middle|X = x\right]$ by fitting a prediction model to the treated arm ($a = 1$) and the control arm ($a = 0$), respectively. The technique is often used in estimating conditional treatment effects (21,35) or identifying subgroups from RCTs (36,37), whereas our focus is counterfactual prediction under interventions. In Alaa *et al.* (26), under a set of assumptions, this technique was adapted for counterfactual prediction with observational data, which used a more complex regression model to address for selection bias in the observational dataset.

Alaa *et al.* (26) adopted standard assumptions of *unconfoundedness* (or *ignorability*) and *overlap (or positivity)*, which is known as the 'potential outcomes model with unconfoundedness'. Their idea is to use the signal-in-white-noise model for the potential outcomes and estimate two target functions, the treated and the untreated, simultaneously with training data. The estimation is done through one loss function, known as the precision in estimating heterogeneous effects (PEHE), which jointly minimises the error of factual outcomes and the posterior counterfactual variance, in such a way to adjust for the bias between the treated and untreated groups. The counterfactual prediction for either treated or untreated can then be made through the estimated posterior mean of two potential outcome functions. Since the ground truth counterfactual outcomes are never available in real-world observational datasets, it is not straightforward to evaluate causal prediction algorithms and compare their performances, a semi-synthetic experimental setup was adopted in (26), where covariates and treatment assignments are real but outcomes are simulated.

For the longitudinal setting where the event history is fully observed, Arjas (27) adopted a marked point process (MPP) framework with a Bayesian non-parametric hazard model to predict the outcome under a single intervention. Point processes are distributions over sequences of time points, and a marked point process is made by attaching a characteristic (a *Mark*) to each point of the process (38). The idea is to incorporate all observed events in the data, including past treatments, covariates and outcome of interest, into a single MPP: $\{(T_n, X_n): n \geq 0\}$, where $T_0 \leq T_1 \leq \cdots$ are the ordered event times and $X_n$ is a description of the event occurring at $T_n$. The model assumed *local independence* – i.e. the intensities of events (that is, the probability of an event occurring in an infinitesimal time interval) when considered relative to the histories $\overline{\mathrm{H}}_t$ are locally independent of outcome risk functions in the model. Under this assumption, in order to define a statistical model for MPP, it suffices to specify the outcome intensities with respect to $\overline{H}_t$ and there is no need for other event time intensities. Prediction under hypothetical interventions can be then made by evaluating the corresponding predictive probabilities in the Bayesian posterior predictive setting given the data.

Both methods in this subsection can be computationally intensive as the number of observed samples increased. This could be ameliorated using conventional sparse approximations (26,39). Both methods are limited to binary interventions, and prediction via treatment effect estimation can only make counterfactual prediction for outcomes with or without intervention.

#### 3.2.2 Counterfactual prediction models that consider time-dependent treatments and treatment-confounder feedback

Papers included in this category (28–33) covered three types of approaches to deal with scenarios where the treatments of interest and confounders vary over time. One example of such confounding is in the sequential-treatment assignment setting, where doctors use a set of variable measurements, at the current time or in the past, to determine whether or not to treat, which in turn affects values of these variables at a subsequent time. For example, whether or not statins are taken at a particular time will affect cholesterol, and these subsequent cholesterol levels affect subsequent decisions about statins. The benefit of such approaches is that they allow consideration of a longer term treatment plan, such as comparing taking statins continuously for ten years from baseline, versus not taking statins for the next ten years. The assumptions needed for identifying unbiased treatment effects in such scenarios are *consistency*, *positivity*, and *sequential ignorability*.

***Marginal structural models (MSMs) within a prediction model framework***

Consider in our running example the hypothetical risk of not taking statins for the next ten years. To account for *treatment drop-ins*, i.e., treatments initiated post-baseline, one straightforward way is to censor patients at treatment initiation; however, this assumes that treatment initiation is non-informative about the baseline and time-dependent covariates. Pajouheshnia *et al*. (8) proposed censoring followed by reweighting using inverse probability of censoring weights (IPCW) to solve the issue of informative censoring in estimating treatment-naïve risk. The proposed method derived time-varying censoring weights by estimating the conditional probabilities of treatment initiation, and then developed a weighted Cox model in the treatment-naïve pseudo-population. A more flexible way of addressing treatment drop-ins for hypothetical risk prediction is to use MSMs with inverse probability treatment weighting (IPTW), where a pseudo-population is created such that treatment selection will be unconfounded. Sperrin *et al.* (28) proposed combining MSM with predictive modelling approaches to adjust for confounding and generate prediction models that could appropriately estimate risk under the required treatment regimens. Following the classic development of IPTW for an MSM, the proposed methods develop two prediction models: a treatment model for computing the probability of receiving post-baseline treatments, and an outcome prediction model fitted with the derived weights and with these post-baseline treatments as well as terms for any interactions between treatment and other predictors included as predictors. By carefully defining the required estimand for the target prediction, the proposed framework could estimate risks under a variety of treatment regimens. In the statin example, this means that one can compare CVD risk under a range of different statin treatment plan, although the focus in the paper was on the 'never takes statins' hypothetical prediction. As with approaches described so far in this category, the model only considered a binary treatment (e.g. statins yes/no). The extension to multiple treatment choices for the proposed method is possible in principal; although, the underlying causal structure and resulted model may become too complex.

Similarly to (28), Lim *et al.* (29) adopted the MSM combined with IPTW approach. Instead of using linear or logistic regression models, they embedded the concept into a deep learning framework and proposed a *Recurrent Marginal Structural Network* (RMSN). The model consisted of (1) a set propensity networks to compute treatment probabilities used for IPTW, and (2) a prediction network used to determine the treatment response for a given set of planned interventions.

The benefit of RMSN is that, it can be configured to have multiple treatment choices and outcomes of different forms (e.g. continuous or discrete) using multi-input/multi-output RNNs. This means, in the statin example, one could consider different doses, and indeed consider alternative treatments as well. Treatment sequences can also be evaluated and no restrictions were imposed on the prediction horizon or number of planned interventions. The use of LSTMs in computing the probabilities required for propensity weighting can also alleviate susceptibility of IPTWs to model misspecification. A drawback is that one needs a rich source of longitudinal data to train the model. Moreover, as in general in deep learning models, they lack a clear interpretation.

***Methods based on balanced representation approach***

Matching approaches such as MSM or RMSN combined with IPTW above adjust for bias in the treatment assignments by creating a pseudo-population where the probability of treatment assignments does not depend on the time-varying confounders. Balanced representation approach, as proposed by Bica *et al.* (30), instead aimed for a representation $\Phi$ of the patient history $\bar{H}_t = (\bar{A}_{t-1}, \bar{X}_t)$ that was not predictive of treatment assignments. That is, in the case of two treatment assignments at time $t$, $P(\Phi(\bar{H}_t)|A_t = 0) = P(\Phi(\bar{H}_t)|A_t = 1)$. It can be shown that, in this way, estimation of counterfactual treatment outcomes is unbiased (40). Bica *et al.* (30) proposed a counterfactual recurrent network (CRN) to achieve balancing representation and estimate unbiased counterfactual outcomes under a planned sequence of treatments (such as statins). CRN improved the closely related RMSN model proposed by Lim *et al.* (29) in a way that overcame the fundamental problem with IPTW, such as the high variance of the weights. As with RMSN, both models required hyperparameter tuning. As the counterfactual outcomes were never observed, hyperparameters in both models were optimised based on the error on the factual outcomes in the validation dataset. As noted by the authors in (30), more work on providing theoretical guarantees for the error on the counterfactuals are required.

***Methods with g-computation for correcting time-varying confounding***

Three papers (31–33) were identified using g-computation to correct time-varying confounding and predicting treatment response curves under the potential outcome framework.

Xu *et al.* (31) developed a Bayesian non-parametric model for estimating conditional treatment response curves under the g-computation formula, and provided posterior inference over the continuous response curves. In the statin example, this means that one can estimate cholesterol or any other continuous outcome of interest under a planned sequence of statin treatments (yes/no). The proposed method modelled the potential outcome using a generalized mixed-effects model combining the baseline progression (with no treatment prescribed), the treatment responses overtime, and noise. The goal was to obtain posterior inference for the treatment response, and predict the potential outcomes given any sequence of treatments conditioned upon past treatments and covariate history. There are two limitations to the model here: (1) it assumes independent baseline progression and treatment response components; (2) treatment response models rely on the additive treatment effects assumption and a careful choice of priors based on clinical details to be decided by domain experts.

Soleimani *et al.* (32) extended the approach in Xu *et al.* (31) in two ways: (1) to continuous-time setting with continuous-valued treatments, and (2) to multivariate outcomes. This means, in the statin example, one could simultaneously predict e.g. risk of CVD, cholesterol and SBT under a range of different statin treatment plans (allowing for different doses assigned at different time points). The model has its ability to capture the dynamic response after the treatment is initiated or discontinued by using linear time-invariant systems. Despite being a more flexible model than (31), this model did not overcome two limitations mentioned above.

Schulam and Saria (33) considered another continuous-time setting where both type and timing of actions may be dependent on the preceding outcome. In the statin example, this means both the statin dose and treatment time (initialisation or discontinuation) depend on the preceding cholesterol level. Here, one needs to predict how a continuous-time trajectory will progress under sequences of actions. The goal was to model action-outcome traces $D \equiv \{t_{ij}, Y_{ij}, a_{ij}\}_{i,j}$: for each individual $i$ and irregularly sampled sequences of actions and outcomes. Schulam and Saria (33) proposed a Counterfactual Gaussian process (CGP) model to model the trajectory and derived an adjusted maximum likelihood objective that learned the CGP from observational traces. The objective was derived by jointly modeling observed actions and outcomes using a marked point process (MPP). The potential outcome query can therefore be answered with the posterior predictive trajectory of the outcome model. A key limitation in this model is that it could not model heterogeneous treatment effects arising from baseline variables.

Counterfactual prediction models in this section using g-formula to correct for time-varying confounding are highly flexible and can be adopted for a variety of clinical settings. However, these methods rely on a set of strong assumptions in both discrete-time and continuous-time settings that are generally not testable; for the latter, Schulam and Saria (33) extended Robin's *Sequential No Unobserved Confounders* assumption to continuous-time case and also assumed *Non-informative Measurement Times*.

## 4. Discussion

In this study, we conducted a methodology scoping review, which has identified two main types of causal predictive modelling (methods that allow for prediction under hypothetical interventions), with the main differences between the methods being the source of data from which the causal effects are estimated. We identified that when the causal effects required for the predictions were fully estimated from the observational data, methods are available for predictions under interventions either at a single time point or varying over time. We have collated current approaches within this field, and highlighted their advantages and limitations in the review.

There are recent studies that have performed a review of methods for causal inference all with different focuses: methods in the analyses of RCTs (41); methods based on graphical models (42) or DAGs (43); methods targeting time-varying confounding (44). Our work differs from these reviews, and, to our knowledge, is the first review to focus on methods enabling predictions under interventions (i.e. counterfactual prediction models). A recent review focused on how time-dependent treatment use should be handled when developing prediction models (8). This clarified the targeted estimand of the clinical prediction model of interest, and consider hypothetical risks under no interventions.

Our search terms, defined from the intersection of prediction modelling filters and causal inference keywords, have been made purposely broad to capture relevant literature, albeit with a high number of false-positives driven by the heterogeneity in language across the fields. This could imply a challenge in devising a potentially more effective search strategy for identifying methodological papers on both fields, a challenge as highlighted in Martin *et al.* (19).

This review has synthesised a range of methods, embedded within both statistical and machine learning frameworks. These methods rely on the availability of the DAG that encodes the relationship between all the relevant variables, and a series of assumptions that make it possible to estimate counterfactual predictions from observational data. Approaches described here cover a wide range of data settings and clinical scenarios. Careful thoughts are needed before adopting these methods, and further challenges and gaps for future research remain, which we will discuss here.

Methods combining information from different sources, such as RCTs combined with observational data, provide a natural way to enable counterfactual predictions; however, challenges remain when combining these two settings. Their objectives are not necessarily complementary, leading to distinct populations included in each study (of possibly very different sample sizes), different sets of covariates being measured, and some potential measurement bias. Therefore, combining observational study with RCTs would need more careful consideration, and a good global guidance may be required. Harrell and Lazzeroni (45) laid out some initial steps one can follow toward an optimal decision making using both RCTs and electronic health record (EHR) data. We also refer the reader to the recent PATH (Predictive Approaches to Treatment effect

Heterogeneity) Statement (46,47), developed to provide guidance for predictive analyses of heterogeneity of treatment effects (HTE) in clinical trials. Predictive HTE analysis aims to express treatment effects in terms of predicted risks, and predict which of 2 or more treatments will be better for a particular individual, which aligns closely with our review aim here. However, as motivated by the limitations in the conventional subgroup analyses in RCTs, predictive HTE analysis has focused on regression-based prediction in randomised trials for treatment effects estimation and subgroup identification. Such techniques can be adapted for the purpose of counterfactual prediction. For example, the predictive modelling used in estimating individualised causal effect in (21,35) was applied for counterfactual prediction in the included paper (30). However, as the primary goal in predictive HTE analyses such as (21,35) is not predicting the counterfactual outcome, we did not include them in our review, which may also be deemed as a limitation of this study.

Another obstacle in combining RCTs with observational study is that, while the estimand for causal inference is clearly defined, the prediction estimand, termed the *predictimand* by Van Geloven *et al.* (5), is often unclear in prediction models. There is an emergence of studies arguing that clearly defining the estimand in prediction is important (28,43). Despite these challenges, and that relatively little work has been done in combining RCTs with patient observational data, it remains an opportunity to explore the interplay of these two areas, as noted in the recent survey by Bica *et al.* (48).

Several key challenges arise in dealing with multiple interventions. The term 'multiple treatments' has been commonly used throughout literature, especially when addressing time-varying treatments. However, the same term may refer to very distinct scenarios in different studies, and greater clarity is necessary. The first and the most often seen scenario, is where multiple values/options are observed for a treatment variable, either at a single time point or over time. Treatments in this setting are indeed 'multivariate treatments'. Many approaches in this review are designated to deal with multivariate treatments (24,30–33), or can in principal be extended to this case (25), (28). However, except for the approach in (24), all methods assume treatment effects from different options to be independent; in (24) interactions between treatment options are modelled through the covariance matrix in the Gaussian process prior. Further methodological development could explore ways to incorporate treatment-treatment interactions into the model.

A second scenario of 'multiple treatments' is where there are interventions on several risk factors, which is substantially more complex, but also more realistic. For example, in clinical settings, one could intervene on different risk factors to prevent CVD, and possible interventions include giving antihypertensive drug or lipid-lowering treatment, lifestyle changing (physical activity, smoking and alcohol drinking), or a combination of them. As these interventions take effect on different parts of the causal structure for the outcome, changes in one factor may affect others, e.g., weight gain after smoking cessation (49). Moreover, each clinical intervention scenario will require its own model for identifying treatment effects from observational data (11). Recent studies on estimating causal effect under multiple interventions have explored methods such as marginal structural cox models (50) and parametric g-formula (51). However, despite its apparent need in clinical practice as in the abovementioned example, there appears to be a lack of models for counterfactual predictions under multiple interventions, and future methodological development is required.

Treatment scenarios addressed so far in this review, both time-fixed and time-varying, are static interventions, i.e. treatment assignment under intervention does not depend on the post baseline covariates. In contrast to the static intervention is the *dynamic treatment strategy*, a rule in which treatments are assigned dynamically as a function of previous treatment and covariate history. Methods such as dynamic MSMs introduced by Orellana *et al.* (52) and independently by Van der Laan and Petersen (53), and variants of structural nested models (SNMs) introduced by Robins (54) were proposed to use observational data to estimate the optimal dynamic treatment regime. Embedding these methods within clinical prediction framework could enable counterfactual predicting under dynamic treatment allocation and support decision-making on optimal treatment rules, which presents a promising avenue for future research (55).

The most pressing problem to address for predictions under hypothetical interventions is model validation. Validation is a crucial step in prediction modelling (counterfactual or otherwise), but is challenging in the counterfactual space since that the counterfactual outcomes are not observable in the validation dataset. The included papers have by-passed this issue by noting that, models are fitted based on the error on the factual outcomes in the validation dataset. In this context, handling of treatment in validation of clinical prediction models has received some attention (56). Pajouheshnia et al. (56) addressed the specific case of validating a prognostic model for treatment-free risk predictions in a validation set where risk-lowering treatments are used. However, further studies are required to extend the potential methods to address the more complex issue in validating counterfactual predictions, such as non-discrete treatment types and non-parametric models as included in this review. While there is emerging research on developing a model validation procedure to estimate the performance of methods for causal effects estimation (57) and sensitivity analysis in causal inference (58), techniques are required to validate the models tailored for counterfactual prediction. Just as domain knowledge is important in causal inference before real-world deployment, it is also important in validating counterfactual prediction, and integrating data generated from RCTs and observational studies and their corresponding models provides a promising way to aid the process (48).

## 5. Conclusions

Prediction under hypothetical intervention is an emerging topic, with most methodological contributions published after 2015. This is now an active area of research in both the statistics and machine learning communities. Available methods for causal predictive modelling can be divided into two approaches. The first combines data from randomised controlled trials with observational data, while the second approach uses observational data only. We recommend using causal effects from randomised controlled trials where possible, combining these with prediction models estimated from observational data, as this alleviates the required assumptions for the causal contrasts to be unbiased. However, further theoretical guarantees are required regarding triangulating data from multiple sources. As well as the data sources available, the targeted estimand needs careful thought, and a relevant approach for the required estimand should be chosen. For example, marginal structural models can be used if observational data are used to make hypothetical predictions concerning an intervention that is sustained into the future. However, techniques to validate such models, and approaches for hypothetical risks under multiple or dynamic intervention scenarios, are under-investigated.

## Declarations

### Ethics approval and consent to participate

Not applicable.

### Consent for publication

Not applicable.

### Availability of data and supporting materials section

Data sharing not applicable to this article as no datasets were generated or analysed during the current study.

### Acknowledgments and funding

This work was funded by the Alan Turing Institute under the 'Predictive Healthcare' project (Health and Medical Sciences Programme). DAJ is funded by the National Institute for Health Research Greater Manchester Patient Safety Translational Research Centre (NIHR Greater Manchester PSTRC. The views expressed are those of the author(s) and not necessarily those of the NHS, the NIHR, or the Department of Health and Social Care.

### Author contributions

All authors contributed to developing the review protocol. LL conducted the literature searches, screening, and data extraction. DAJ conducted initial 3% abstract screen to ensure reliability of the screening process. MS contributed to the study selection against inclusion/exclusion criteria. LL and MS wrote the first draft. All authors discussed, reviewed and edited the manuscript, and have approved the final version.

### Declaration of conflicts of interest

The authors declare that they have no conflicts of interest relating to the publication of this work.